# Design of GaN White Light Emitting Diode through Envelope Function Analysis and Combined *k.p*-Transfer Matrix Method

Milad Khoshnegar, Majid Sodagar, Amin Eftekharian, Sina Khorasani

School of Electrical Engineering, Sharif University of Technology, P. O. Box 11365-9363, Tehran, Iran

**Abstract**
In this paper, we present an envelope function analysis in order to design the emission spectra of a white quantum well light emitting diode. The nanometric heterostructure that we are dealing with is a multiple quantum well, consisting periods of three single quantum wells with various well thicknesses. With the aid of 6×6 Luttinger Hamiltonian, we employ the combination of two methods, *k.p* perturbation and transfer matrix method, to acquire electron and hole wavefunctions analytically. The envelope function approximation was considered to obtain these wavefunctions for a special basis set. While adjacent valence subbands have been studied exactly, the conduction bands are approximated as parabolic. The effect of Stokes shift has been also taken into account. The dipole moment matrix elements for interband atomic transitions are evaluated via correlation between electron and hole envelope functions, for both orthogonal polarizations. This has simplified the calculation of photoluminescence intensity. Spatial variations in hole/electron wavefunctions have been examined with the introduction of piezoelectric and spontaneous polarizations internal field. We theoretically establish the possibility of a highly efficient InGaN red emitter, resulting in a uniform luminescence in red, green and blue emissions from the while light emitting diode, through adjusting material composition, potential slope, and thickness.

## 1. Introduction

Recently, wurtzite strained quantum wells have been studied intensively in diode laser and RGB Light Emitting Diode (LED) structures. In this regard, group III nitride-based QWs have earned an extensive contribution in multicolor LEDs. Especially, white LEDs have attracted a great attention in the field of solid state lighting. To realize white Quantum Well LEDs (QWLEDs), two major schemes are normally proposed. The first one is using a converting layer, conventionally phosphorous, to obtain a redshift from short wavelengths [1-3]. This method has several disadvantageous like low efficiency, complex packaging and short lifetime due to degeneration of phosphor material. Also, Stokes shift energy loss is inevitable in these kinds of white LEDs [4]. The second scheme is based on exploiting several QWs, each emitting in a definite wavelength, and combining the output photoluminescence spectra.

Until now, white-light LEDs, including InGaN/GaN QWs, which simultaneously emit two or three colors, are reported [5-9]. All of the reported structures, however, suffer from low internal quantum efficiency and low intensity emission of the red wavelength, which is principally due to the lack of an efficient red emitter. As a result, with a lower spectral peak for red QW, the photoluminescence spectrum is reduced for long wavelengths, leading to a nonuniform spectrum of white light output. Moreover, efficient InGaN/GaN LEDs within the range $\lambda = 630 \sim 640\,nm$ are nearly hard to access. Most attempts have therefore utilized yellow emitters to obtain long wavelength [10-11]. This is possibly due to the required composition of well/barrier: In order to narrow the band gap for red emission a high Indium contribution within InGaN is inevitable, which yields nearly strong internal field and Stark effect that may not provide the appropriate overlap between electron/hole distributions within emitters.

Here, we intend to give an analytical study on GaN-based LEDs, in order to realize the effect of physical parameters on optical intensity spectrum. For this purpose, the QW heterostructure is basically analyzed by its band structure and wavefunctions. The first point in this case is that the group III nitrides have a wurtzite crystal structure. The near band edge Hamiltonian of these structures have been extracted dependent upon the symmetry properties of crystal [12], while band curve parameters have been obtained empirically. The theoretical parameters in this case, however, have shown some inconsistency with experimental data. On the other hand, one can rely on *k.p* method to derive Hamiltonian matrix for GaN, InN, AlN and their alloys [13]. Although *k.p* is able to give bulk band structures, for analysis of multilayer hetero-structures other methods must be supplemented to fully describe the system behavior. In the literature two main approaches are introduced: finite difference [14] and transfer matrix [15] methods. The former is suitable for multi-dimensional systems such as quantum dots. Here we utilized the latter, because of structural simplicity.

The main goal of analytically studying the emission region in a QWLED is interpretation of interband atomic transitions. With the aid of the Hamiltonian given by Chaung *et al.* [14], we could find both the band structure and electron/hole wavefunctions within the QW. Having the eigenwaves known, calculation of dipole moment matrix elements for interband transitions became straightforward. In the next step, we have calculated optical gain (intensity) and studied structural parameters. An important phenomenon in wurtzite QWs is strain. By neglection of internal field, the strain itself can cause major changes in band structure and hence intensity spectrum. Taking the strain-dependent piezoelectric polarization field into account, this effect becomes even more dominant. Therefore adjusting well/barrier alloy composition is very influential on output spectra. We have shown that it would be possible to construct three QWs with an output spectrum containing equal spectral peaks belonging to the three basic red, green, and blue wavelengths. Chromaticity coordinates have been constructed and it is shown that a true white emission is expected.

## 2. Combined *k.p* and Transfer Matrix Method

In contrast to bulk materials, in low dimensional structure (i.e., quantum wells, wires and dots), the potential experienced by electrons is no longer periodic in all directions and hence the exact eigenstates corresponding to such structures may not be obtained readily. Here, we employ envelope function approximation (EFA) in order to simplify the problem. In this method, the bulk band structure is considered as a macroscopic potential modulated by microscopic atomic potentials. The validity of such approximation depends on the nature of bands. According to variational method, EFA is justified where the band under consideration is not degenerate and further the energy distance between neighboring bands is not so small. In case of conduction band (CB) corresponding to zinc blend, diamond or wurtzite structure semiconductors, employing single band EFA is sufficient for obtaining eigenstates. That is why it is non-degenerate (excluding spin effect), and its energy is far from other bands.

Here, the atomic S-like orbital can be considered as the Bloch part of total wave function. In contrast, valence band in such structures is two-fold degenerate at Γ point due to heavy hole (HH) and light hole (LH) bands. Furthermore, the spin-orbit split-off (SO) band is not far away. In this case, the effect of these neighboring bands on hole eigenstates is not negligible and should be taken into account. Although one can exploit the effect of other neighboring bands such as CB bands, the results does not change significantly. As a result, the Bloch part for total wave function cannot be recognized. Instead, a combination of several envelope and Bloch functions constitutes the total wave function. In order to model the governing Hamiltonian corresponding to holes, we employ Luttinger approach, which results in several envelop functions corresponding to particular Bloch waves. In this approach, the total wave function of the state $|\nu\rangle$ can be written as

$$\langle x|\nu\rangle = \sum_i \psi_i u_i \qquad (1)$$

where, $\psi_i$ and $u_i$ stand for *i*th envelope function and its corresponding Bloch part, respectively. The upper bound for $i$ is determined by the number of bands taken into account. By considering three bands, i.e. HH, LH and SO, Luttinger Hamiltonian matrix representation in the basis set consists of $\pi$ orbitals, i.e. $|x\uparrow\rangle$, $|x\downarrow\rangle$, $|y\uparrow\rangle$, $|y\downarrow\rangle$, $|z\uparrow\rangle$ and $|z\downarrow\rangle$, results in a 6×6 matrix [16]. It is a common practice to block-diagonalize it by performing a unitary transformation over basis set. It should be kept in mind that upon transforming original basis set, the envelope functions no longer remain symmetric. That is why the elements of the resultant basis set are not symmetric themselves. After employing such unitary transformation, the resultant matrix reads [14]

$$\mathbf{H}_{6\times 6} = \begin{pmatrix} \mathbf{H}^U & \mathbf{O} \\ \mathbf{O} & \mathbf{H}^L \end{pmatrix} \qquad (2)$$

where the upper and lower blocks are defined through

$$\mathbf{H}^{\sigma=U,L} = \begin{pmatrix} F+V & K_P & \mp iH_P \\ K_P & G+V & \Delta \mp iH_P \\ \pm iH_P & \Delta \pm iH_P & \lambda + V \end{pmatrix} \qquad (3)$$

in which $V$ stands for potential profile that could have contribution from band offset, spontaneous polarization field, piezoelectric field or external electric field. The rest of parameters are defined as [14]:

$$F = \Delta_{cr} + \Delta_{so}/3 + \lambda + \theta \qquad (4a)$$

$$G = \Delta_{cr} - \Delta_{so}/3 + \lambda + \theta \qquad (4b)$$

$$\lambda = -\frac{\hbar^2}{2m_0}\left[(\gamma_{1z} + 4\gamma_{3z})k_z^2 + (\gamma_{1P} - 2\gamma_{3P})k_P^2\right] + \lambda_\varepsilon \qquad (4c)$$

$$\lambda_\varepsilon = -(\delta_{1z} + 3\delta_{3z})\varepsilon_{zz} - (\delta_{1P} - 2\delta_{3P})(\varepsilon_{xx} + \varepsilon_{yy}) \qquad (4d)$$

$$\theta = \frac{\hbar^2}{2m_0}(6\gamma_{3z}k_z^2 - 3\gamma_{3P}k_P^2) + \theta_\varepsilon \qquad (4e)$$

$$\theta_\varepsilon = 6\delta_{3z}\varepsilon_{zz} - 3\delta_{3P}(\varepsilon_{xx} + \varepsilon_{yy}) \qquad (4f)$$

$$K_P = -\frac{\hbar^2}{2m_0}(\gamma_{2P} + 2\gamma_{3P})k_P^2 \qquad (4g)$$

$$H_P = -\frac{\hbar^2}{2m_0}\sqrt{2}\left(2\gamma_{2z}+\gamma_{3z}\right)k_z k_P \tag{4h}$$

$$\Delta = \sqrt{2}/3\Delta_{so} \tag{4i}$$

In case of QW, one can simply assume the in plane wave vector $k_P = \sqrt{k_x^2+k_y^2}$ is a scalar and $k_z$ stands for $-i\partial/\partial z$ operator. Also, $\varepsilon_{xx}$, $\varepsilon_{yy}$, and $\varepsilon_{zz}$ are the strain components due to lattice mismatch. Here we assume that the semiconductor is grown on (0001) orientation; consequently, other strain components, i.e. $\varepsilon_{xy}$, $\varepsilon_{yz}$, and $\varepsilon_{zx}$ vanish. $\gamma_{iP}$ and $\gamma_{iz}$ denote in and out of plane Luttinger parameters, respectively. These parameters vary across the barrier-well interfaces and can be determined experimentally. Also, $\delta$, $\Delta_{so}$ and $\Delta_{cr}$ denote deformation potential, split-off energy and the crystal field splitting energy, respectively. By employing quasi-cubic approximation which is justified in wurtzite structures, one can simply assume [17]

$$\gamma_{iz} \approx \gamma_{iP}, \quad i \in \{1,2,3\} \tag{5a}$$
$$\delta_{1z} \approx \delta_{1P} = V_a \tag{5b}$$
$$\delta_{2z} \approx \delta_{2P} \approx \delta_{3z} \approx \delta_{3P} \approx -V_b/2 \tag{5c}$$

where $V_a$ and $V_b$ are hydrostatic and shear deformation potentials. Notice that the above-mentioned unitary transformation depends on in plane wave vector. The exact transformation can be found elsewhere [18,19].

In general, the eigenvalue problem corresponding to the above Hamiltonian can be solved through geometry-independent finite difference (FD) method. However, QW structures can be treated more readily owing to their simple layered structure. In this work, transfer matrix method (TMM) is employed. Note that the upper and lower blocks in (2) are related through the operator $\mathbf{H}^U = (\mathbf{H}^L)^*$. Therefore, the eigenvectors of $\mathbf{H}^L$, can be achieved upon determining that of $\mathbf{H}^U$ through complex conjugate transformation. Henceforth, we will focus on upper block eigenvalue problem, i.e. $\mathbf{H}^U \psi = \zeta \psi$. It is straightforward to decompose $\mathbf{H}^U$ in three terms with respect to $k_z$ [16]:

$$\mathbf{H}^U = \mathbf{H}_2 k_z^2 + \mathbf{H}_1 k_z + \mathbf{H}_0 \tag{6}$$

If we assume constant potential profile in each layer, then all $k_z$ coefficients will be constant matrices. Although, in practical cases where potential profile is not constant, one can simply assume staircase approximation and choose stair width small enough so that the corresponding error becomes negligible. The eigenvalue problem results in three coupled second order differential equations in terms of $k_z$. It is possible to reduce the differential equation order by introducing $\mathbf{\Phi} = [\psi, \psi']^T$ at the cost of increasing the number of coupled equations to six. If so, the governing equation for $\mathbf{\Phi}$ reads [20]

$$\mathbf{\Phi}' = \mathbf{\Lambda}\mathbf{\Phi} \tag{7a}$$

$$\mathbf{\Lambda} = \begin{pmatrix} \mathbf{O} & \mathbf{I} \\ \mathbf{H}_2^{-1}(\mathbf{H}_0 - \zeta\mathbf{I}) & -i\mathbf{H}_2^{-1}\mathbf{H}_1 \end{pmatrix} \tag{7b}$$

Note that $\mathbf{\Lambda}$ is not diagonal and hence (7a) cannot be solved directly. It can be decomposed as

$$\mathbf{\Lambda} = \mathbf{PDP^{-1}} \tag{8}$$

Here, $\mathbf{D}$ is a diagonal matrix composed of $\mathbf{\Lambda}$ eigenvalues. Also, $\mathbf{P}$ is a square matrix composed of eigenvectors related to $\mathbf{\Lambda}$. Notice that this decomposition is not unique. We may choose $\mathbf{D}$ so that its upper and lower half contain elements with positive and negative real part, i.e. forward and backward propagating waves, respectively. By introducing another change of variable, as $\mathbf{Q} = \mathbf{P^{-1}\Phi}$, (7a) recasts into $\mathbf{Q}' = \mathbf{DQ}$ with the solution

$$\mathbf{Q} = e^{\mathbf{D}}\mathbf{Q_0} \tag{9}$$

It is evident that envelope wave function and probability current continuity boundary conditions across an interface should be imposed on $\mathbf{\Phi}$, rather than $\mathbf{Q}$. Assuming that Bloch phases across interfaces remain unchanged, it is straightforward to write down

$$\mathbf{B}_L\mathbf{\Phi}(z_0^L) = \mathbf{B}_R\mathbf{\Phi}(z_0^R) \tag{10}$$

where the boundary condition matrix $\mathbf{B}$ is [20]

$$\mathbf{B} = \begin{pmatrix} \mathbf{I} & \mathbf{O} \\ -i\mathbf{H}_1 & -\mathbf{H}_2 \end{pmatrix} \tag{11}$$

The total transfer matrix can be constructed by multiplying all transfer matrices correspond to each layer and applying appropriate boundary conditions across each interface. This would result in

$$\mathbf{T} = \prod_{i=1}^{l-1} \mathbf{P}_{i+1}^{-1}\mathbf{B}_{i+1}^{-1}\mathbf{B}_i\mathbf{P}_i e^{\mathbf{D}_i \Delta z_i} \tag{12}$$

where $\mathbf{P}_i$, $\mathbf{B}_i$, $\mathbf{D}_i$, and $\Delta_{zi}$ are the eigenvectors matrix, boundary condition matrix, eigenvalues matrix and length of $i$th layer, respectively. Also, $l$ is the number of layers. As it can be traced, the total transfer matrix will depend on system energy $\zeta$. In order to determine the permitted energies, wave function normalization should be considered. If $a_i$ and $b_i$ denote forward and backward wave vectors in $i$th layer, respectively, the following relation should hold for bound states when $a_1$ is set to zero [18]

$$a_l = \mathbf{T}_{11}(\zeta)a_1 + \mathbf{T}_{12}(\zeta)b_1 = \mathbf{T}_{12}(\zeta)b_1 \tag{13a}$$
$$0 = \mathbf{T}_{21}(\zeta)a_1 + \mathbf{T}_{22}(\zeta)b_1 = \mathbf{T}_{22}(\zeta)b_1 \tag{13b}$$

Equation (13b) shows that a particular $\zeta$ is the system eigenenergy, or equivalently the system has non-trivial bound state, if and only if $\mathbf{T}_{22}$ has an eigenvalue equal to zero for that specific $\zeta$. In addition, the corresponding eigenvector should be scaled so that the resulting wave function remains normalized. In order to find acceptable $\zeta$, the system energy can be swept while the determinant of $\mathbf{T}_{22}$ is monitored. The only difficulty of the above approach is the appearance of spurious states [20]. This is probable because such procedure is perturbative, in which an incomplete basis set is used. These spurious states reveal themselves in rather large eigenvalues of $\Lambda$ for some layers and hence may result in instability. This problem can be treated by eliminating such eigenvalues in $\mathbf{D}$ matrix or clamping them appropriately [15]. In subsequent sections, we will use this procedure to achieve our QW envelope functions.

## 3. Wavefunctions and Optical Gain

Now, we may proceed to calculate dipole moments related to interband transitions between conduction and valence bands. After obtaining six envelope functions via $k.p$-TMM, we can readily define hole wavefunction of $\vartheta$ th valence band as [21]

$$\Psi^{v,U}_{\vartheta,k_P} = \mu^U_\vartheta e^{i\mathbf{k}_P \cdot \mathbf{r}_P} \left( \psi^{U,1}_\vartheta |v^U_1\rangle + \psi^{U,2}_\vartheta |v^U_2\rangle + \psi^{U,3}_\vartheta |v^U_3\rangle \right) \quad (14a)$$

$$\Psi^{v,L}_{\vartheta,k_P} = \mu^L_\vartheta e^{i\mathbf{k}_P \cdot \mathbf{r}_P} \left( \psi^{L,1}_\vartheta |v^L_1\rangle + \psi^{L,2}_\vartheta |v^L_2\rangle + \psi^{L,3}_\vartheta |v^L_3\rangle \right) \quad (14b)$$

where $|v_i^{\sigma=U,L}\rangle$ were explained in the pervious section. For either $\sigma = U$ or $L$, the aforementioned wavefunctions must be normalized; therefore, we have represented normalization coefficient with $\mu^\sigma_\vartheta$. On the other hand, near the bottom of conduction bands, we can consider S-like symmetry for eigenstates. In contrast with $\Psi^{v,\sigma=U,L}_{\vartheta,k_P}(k_p)$, wavefunction independency of in-plane wavevector is acceptable and the parabolic approximation for band structure is justified. Hence, we get

$$\zeta^c_o(k_P) \approx \zeta^c_o(k_P = 0) + \frac{\hbar^2 k_P^2}{2 m_e^P} \quad (15a)$$

$$\Psi^{c\eta}_{o,k_P}(z) = \mu_o e^{i\mathbf{k}_P \cdot \mathbf{r}_P} \phi_o(z) |S,\eta\rangle \quad (15b)$$

in which, $m_e^P$ is the in-plane component of electron effective mass, $\Psi^{c\eta}_{o,k_P}$ denotes wavefunction of electron in $o$th conduction band, $\phi_o(z)$ stands for electron envelope function and finally $\eta$ is its spin direction along $z$ axis.

To analyze the optical intensity spectrum when only interband transitions are taken into account, dipole moments should be derived. For this purpose, if we consider $M^{\eta,\sigma}_{o\vartheta}(k_P)$ as the transition matrix element of dipole element from the state with the wavefunction of $\Psi^{c\eta}_{o,k_P}$ to the state with the wavefunction of $\Psi^{v,\sigma}_{\vartheta,k_p}$, then the optical gain for the two polarizations is defined as ($e$=0,1 stand for TE, TM polarizations, respectively)

$$g^e_{sp}(\hbar\omega) = \frac{q^2}{4\pi n_r c \varepsilon_0 m_0^2 \omega L_\omega} \sum_{\eta=\uparrow,\downarrow} \sum_{\sigma=U,L} \sum_{o,\vartheta}$$
$$\times \iint \left| \hat{e} . M^{\eta,\sigma}_{o\vartheta}(k_P) \right|^2 \Xi(\omega,k_P) k_P dk_P d\varphi \quad (16)$$

Here, $n_r$ is the well refractive index, $q$ and $m_0$ are the free electron charge and mass, respectively, $L_\omega$ is the length of well region, and $\hat{e}$ denotes the electric field's polarization vector. Furthermore, $\Xi(\omega,k_P)$ is strongly dependent on Fermi distributions in both valence and conduction bands and also on the position of quasi Fermi levels [21]:

$$\Xi^\sigma_{o,\vartheta}(\omega,k_P) = \frac{f^c_o(k_P)\left(1-f^v_{\sigma\vartheta}(k_P)\right)(\hbar/\tau\pi)}{\left(\zeta^c_o(k_P)-\zeta^v_{\sigma,\vartheta}(k_P)-\hbar\omega\right)^2 + (\hbar/\tau)^2} \quad (17)$$

where, $f^c$ and $f^v$ are electron and hole Fermi distribution functions, respectively. We can adjust interband relaxation time $\tau$ to tune the half linewidth of Lorentzian function $1/\tau$, that is commonly in the order of $(PS)^{-1}$ for zinc-blend or wurtzite crystals [22]. With respect to QW band structure, in order to determine carrier density, one can alter quasi Fermi levels. This automatically affects Fermi distributions and hence the $\Xi$'s amplitude. For a definite quasi Fermi level in conduction band $F_c$, the electron concentration can be achieved by [23]

$$n = \frac{1}{\pi L_\omega} \sum_o \int f^c_o(k_P) k_P dk_P = \frac{k_B T m_e^P}{\pi \hbar^2 L_\omega}$$
$$\times \sum_o \ln\left[1 + e^{[F_c - \zeta^c_o(0)]/K_B T}\right] \quad (18)$$

Similarly, hole concentration is obtained from

$$p = \frac{1}{2\pi L_\omega} \sum_{\sigma=U,L} \sum_\vartheta \int \left[1 - f^v_{\sigma\vartheta}(k_P)\right] k_P dk_P \quad (19)$$

According to (16) and (17), it is noteworthy to mention that although several eigenstates for both conduction and valence bands were taken into account, the behavior of optical gain spectrum is chiefly affected by subbands close to bottom of conduction band and top of valence band. This is strongly due to quasi Fermi levels and the impact of Fermi distributions on gain magnitude.

To distinguish the term $\mathcal{G}^{\sigma,\hat{e}}_{o\vartheta} = \left|\hat{e}.M^{\eta,\sigma}_{o\vartheta}(k_P)\right|^2$ in (16), we follow the approach introduced in [21]. In this reference, it is shown that after some simplifications, the efficient expression that can be applied in (16) instead of $\mathcal{G}^{\sigma,\hat{e}}_{o\vartheta}$ for TE polarization is

$$\mathcal{G}_{o,\vartheta}^{\sigma,\hat{x}} = \frac{1}{4}\left|\langle S|p_x|X\rangle\right|^2 \left\{ \langle \phi_o|\psi_\vartheta^{\sigma,1}\rangle^2 + \langle \phi_o|\psi_\vartheta^{\sigma,2}\rangle^2 \right\} \quad (20a)$$

while for TM polarization takes the form

$$\mathcal{G}_{o,\vartheta}^{\sigma,\hat{z}} = \frac{1}{4}\left|\langle S|p_z|Z\rangle\right|^2 \langle \phi_o|\psi_\vartheta^{\sigma,3}\rangle^2 \quad (20b)$$

for both $\eta = \uparrow$ and $\eta = \downarrow$. In addition to the correlation terms above, we need to evaluate $\left|\langle S|p_x|X\rangle\right|^2$ and $\left|\langle S|p_z|Z\rangle\right|^2$ by utilizing Kane's model for wurtzite crystals and considering hexagonal symmetry [21]. Then these parameters are definable through well energy gap and split-off energy. Finally, (16) is simplified as

$$g_{sp}^e = \frac{q^2}{n_r c \varepsilon_0 m_0^2 \omega L_\omega} \sum_{\sigma=U,L} \sum_{o,\vartheta} \int \mathcal{G}_{o,\vartheta}^{\sigma,\hat{e}} \Xi_{o,\vartheta}^\sigma (\omega, k_P) k_P dk_P \quad (21)$$

Then the material gain reads

$$g_m = g_{sp}^e \left(1 - \exp\left[(\hbar\omega - F_c - F_v)/K_B T\right]\right) \quad (22)$$

It is noteworthy to remind that above multiband calculations seems unavoidable if one is dealing with position dependent potentials, e.g. quantum wells or dots. In the bulk material, the pure state corresponding to a definite single band may exist. However, in structures such as QWs, an eigenstate is a mixture of pure states and all bulk bands contribute in its wavefunction. The multiband analysis aids us to identify the involved transitions and wavefunctions in the photoluminescence intensity of polarized light; and this is a result of accurate description of dipole moment matrix elements in atomic transitions.

## 4. Internal Field and Gain Spectrum

While dealing with the Hamiltonian (3), we need to study the biaxial strain further than other physical parameters. Biaxial strain is tunable through well/barrier mismatch. However, it does not appear for all thickness ratios of well and barrier: it is shown that for some critical Indium fractions in $In_xGa_{1-x}N/GaN$ QWs, and for some proportions of W/B thicknesses, strain relaxation occurs in the form of pinholes [24].

From the band structure viewpoint, generally, when in-plane strain is negative, energy gap becomes smaller and vice versa. In addition, biaxial strain has a serious impact on piezoelectric polarization in wurtzite QWs

$$P_{pz} = d_{31}\left(C_{11} + C_{12} - 2\frac{C_{13}^2}{C_{33}}\right)\varepsilon_{xx} \quad (23)$$

Here, $d_{31}$ is a part of piezoelectric tensor which itself varies with $\varepsilon_{xx}$ linearly [24]. This polarization beside spontaneous polarization makes a slanted profile for either well or barrier potential. A slanted potential has two major influences on intensity spectrum. First is the relocation of adjacent subbands and hence changing spectrum's peak in a fixed range of wavelength. The other effect is drift of electron and hole envelope functions along the perpendicular $z$ axis, which affects correlation terms in $\mathcal{G}_{o,\vartheta}^\sigma$; therefore it modifies peak amplitude by varying overlaps of electron/hole distributions. On the other hand, spontaneous polarization is free of strain. So, its internal field can be intensified or canceled out with that of piezoelectric polarization to make appropriate spectrum redshift or blueshift.

In addition to the biaxial strain, one can rely on well length to determine subbands distances. For a fixed value of well/barrier length and in the absence of internal field, enlarging thickness of well leads to nearing valence subbands, especially at $k_P=0$ [14]. Then, for a definite carrier concentration, more valence subbands contribute in optical intensity spectrum. However, when the quantum confined Stark effect (QCSE) is considered, well depth is principally influential: for a definite internal field slope $E_i$, the larger the well depth $E_i \times L_W$, the narrower the energy gap; this brings about a redshift in the spectrum and a reduction in electron/hole wavefunction overlap.

To analyze envelope functions in the presence of internal field, one needs to insert a position varying potential in Hamiltonian (3). Due to the fact that finding eigenvalues in this case is demanding, we approximated potential with a series of step-like profiles. Then, the entire system was modeled with numerous constant potentials and owing to TMM, we could obtain envelope functions readily. The LAPACK routines were utilized to calculate transfer matrices and search for zero determinants.

## 5. Exciton Localization and Stokes' Shift

Currently, growth of high-quality $In_xGa_{1-x}N$ films has been made possible with the aid of molecular beam epitaxy, even with high Indium index [25]. During studying these alloy samples with such high Indium content, a relatively serious difference is observed between emission and absorption spectral peaks, referred to as the Stokes' shift. As the low miscibility of InN in GaN yields great composition fluctuations of indium or phase segregation within InGaN layers [26], some articles attribute the photoluminescence emission to the localized excitons at In-rich quantum dots which ordinarily cause lower band gaps [27]. In this model, electrons and holes excited by photons accumulate in clusters and increase the overlap of electron/hole distributions, leading an enhancement in optical intensity. A relevant simulation by C. S. Xia *et al.*

reveals that taking this model into account has no significant role in alteration of spectral profile, but it rather contribute to the position of spectral peaks [28]. Therefore, we considered excitonic localization effect as a Stokes'-like shift in emitters' spectra. It is shown experimentally that the redshift increases linearly with Indium contribution within $In_xGa_{1-x}N$ for $x<0.5$ [29]. In practice, this band gap narrowing caused by composition inhomogeneity inside InGaN layers aids us reaching red emitter with lower indium indices, if we choose InGaN as the well material.

## 6. Proposed Structure

At present, there have been several geometries suggested for growing on-chip multi-emitters to give a relatively monolithic white light. Although the attitude of new trends is to propose novel structures such as trapezoidal InGaN/GaN MQWs [30], the main contribution is for the simple laterally [5] or vertically [31] distributed MQWs. With constructing the *p-n* junction in the transverse direction, the lateral injection current causes a nearly uniform carrier concentration for all emitters. The nonuniform carrier density, which may be attributed to accumulation of injected carriers in the QWs near *p*-GaN, is a common problem in some common devices [7] and is almost suppressed in vertical configuration [31].

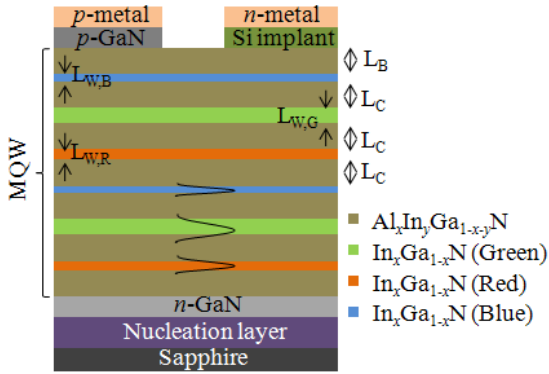

Fig. 1. Schematic of a simple MQW containing three emitters grown vertically within each period is shown (only two periods are depicted here and MQW is enlarged for clarity). $L_C$, spacing between two adjacent single QWs, must be adjusted to minimize overlaps of corresponding envelop functions, e.g. $L_C>L_B$.

In Fig. 1 we have depicted the schematics of the proposed vertically grown MQW, composed of three emitters within each period, between *p*-GaN and *n*-GaN layers. The well and barrier compositions are specified to be $In_xGa_{1-x}N$ and $Al_xIn_yGa_{1-x-y}N$. In the next section, we will explain why we have chosen these well/barrier compositions. $L_{W,G}$, $L_{W,R}$ and $L_{W,B}$ denote the well thickness of green, red and blue emitters, respectively. $L_B$ is the barrier length and $L_C$ is the separation between adjacent QWs. $L_B$ can be exactly determined and may be the lower bound for $L_C$: In MQW design, as the electron/hole envelope functions decay within a few nanometers inside barriers, $L_C$ must be determined so that the wavefunctions from adjacent single QWs have no significant overlap.

## 7. QW Design for White LED

In order to design an on-chip white LED, first we are concerned with the composition of well and barriers. As the consequent photoluminescence spectrum includes all wavelengths in RGB range, the difference between well energy gap in red QW and that of blue QW must be at least 0.8-0.9eV; this is the minimal difference among blue and red spectrum's peak. In addition, well material must be chosen to be able to reach energy band gaps in the limit of 1.9-2.8eV. For GaN/(Al,Ga)N QW this is not achievable, except with a relatively large amount of internal field and strong Stark shift [33], due to the fact that GaN energy band gap is too high for red emission. Therefore, we used $In_xGa_{1-x}N$ as the well composition. For InGaN we used linear interpolation for band parameters, except for energy gap that is [24]:

$$E_g^{InGaN}(x) = xE_g^{InN} + (1-x)E_g^{GaN} - x\beta(x)(1-x) \quad (24)$$

in which $\beta(x \leq 0.2) = 3.8$, $\beta(0.2 < x \leq 0.3) = 2.4$ and $\beta(0.3 < x \leq 0.5) = 1.4$.

In order to recognize the suitable barrier composition we must consider the Stark shift and if the QW depth, $E_i \times L_W$, is comparable to $\Delta E_g$, where $\Delta E_g$ is the difference of well and barrier energy band gaps. In the case that QW depth nears or exceeds the $\Delta E_g$ overlap between electrons, and hole wavefunctions are decreased and optical intensity falls down intensively. Therefore, we selected quaternary composition $Al_xIn_yGa_{1-x-y}N$ for barrier, yielding a relatively large $\Delta E_g$; $E_g(AlN)=6.28eV$. For $Al_xIn_yGa_{1-x-y}N$ we utilized cubic approximation for interpolation, then energy gap and lattice parameters were evaluated with regard to [34]. To determine elastic parameters, we followed Wright *et al.* values for GaN, InN and AlN [35]. Important band parameters are listed in the Appendix.

The photoluminescence spectrum of MQW of Fig. 1, must at least contain three spectral peaks within visible range of wavelength. We have so adjusted the well/barrier compositions, that the field due to piezoelectric polarization cancels out that of spontaneous polarization and the internal field inside green emitter becomes trivial, giving a closely flat potential. Then, with stronger and weaker Stark effects red and blue spectral peaks are achievable. For this purpose, a 16*nm* $Al_{0.24}In_{0.38}Ga_{0.38}N$ and 3*nm* $In_{0.25}Ga_{0.74}N$ barrier and well were chosen, respectively. The biaxial strain and total internal field are estimated to be 1% and 0.35MV/cm respectively, giving an energy band gap equal to 2.58eV, $\zeta_{HH_1} = 0.01\,eV$ and $\zeta_{C_1} = 2.59\,eV$; see Fig. 2 (center). The Stokes' redshift is approximately 0.25eV for $In_xGa_{1-x}N$ and $x=0.25$ [29], giving 2.33eV effective band gap when the excitonic localization within InGaN layer is taken into account. QW depth is 0.092eV which much smaller in compare with $\Delta E_g=0.705eV$, and hence no significant separation occurs in electron/hole distributions. It is remarkable that from this step, the ternary QWs compositions must be selected in a manner that in-plane strain does not exceed e.g. 2%; larger biaxial strains may result in strain relaxation.

For the red emission, well composition was chosen to be $In_{0.32}Ga_{0.68}N$. Higher Indium fraction in $In_xGa_{1-x}N$ well composition has been earlier reported by Chen et al. [36]. The biaxial strain and total internal field was estimated to be 0.2% and −0.9MV/cm and internal field due to spontaneous polarization became dominant. The well thickness is set 2nm to give rise (fall) into first heavy hole band (first conduction band). The QW depth, equal to 0.18eV, is greater than that of green emitter but still much smaller than $\Delta E_g=0.73eV$. This comparison indicates that although the higher Stark effect in red QWs yields a band gap narrowing, the overlaps of electron/hole distributions are not significantly separated and an efficient red emitter is still expected. $\zeta_{HH_1}$ and $\zeta_{C_1}$ were obtained 0.13eV and 2.39eV, yielding a 2.26eV energy band gap, which is appropriate for red emission when a 0.32eV Stokes' redshift by localized excitons is regarded; effective band gap $E_g$ is 1.94eV and $\lambda \approx 641nm$.

In Fig. 2 (left) the valence band structure corresponding to red single QWs is depicted. The separation between valence subbands is mainly contributed by well thickness and internal field. However when the well thickness is still small relatively (2nm) and Stark effect is weak, the band separation follows the impact of well thickness: reducing $L_W$ from 3nm, in green QW, to 2nm, in red QW, results in a subband-stretching and pushing $HH_2$, $LH_2$, and especially $CH_1$ into farther energies from top of the valence band. It is known that envelope functions due to CH subbands chiefly participate in TM optical intensity. When we concern with spontaneous regime, the quasi Fermi level is normally fixed near top energy eigenvalues, e.g. $HH_1$, in order to avoid degenerate condition. As the subbands close to quasi Fermi level, principally determine the optical intensity, having distant $HH_1$ and $CH_1$ causes suppression in TM gain.

Finally, in order to design a blue emitting QW, $In_{0.23}Ga_{0.77}N$ was chosen as the 1nm well composition. In fact, reaching lower wavelength demands for less Indium content, which itself increases biaxial strain and total internal field with respect to green QW. The biaxial strain and $E_i$ rose to 1.17% and 0.84MV/cm. To compensate the probable band gap narrowing, we lessened the well thickness, $L_W=1nm$. Notice that QW depth 0.084eV is still trivial comparing with $\Delta E_g=0.65eV$. In comparison, the band structures given in Fig. 2 (right) and (left) show that enlarging Indium index and reducing well thickness, in an approximately equal internal field modulus, has raised the $CH_1$ energy eigenvalue upper than $HH_2$ and $LH_2$. Actually, during the simulations we observed that $HH_2$ energy eigenvalue was obtained lower than −0.3eV. This subband replacement and separation has two major effects: 1) in blue emission $CH_1$ nears top of valence band and higher TM-polarized photoluminescence intensity is expected relative to red emission. 2) $HH_2$ and $LH_2$, which contribute in TE optical intensity, are far enough that the related transitions have no major effect on TE photoluminescence. Hence, no significant broadening will occur in blue TE-polarized spectral peak. We obtained $\zeta_{HH_1} = -0.033eV$ and $\zeta_{C_1} = 2.82eV$, then imposed 0.23eV Stokes' shift to reach the required effective energy band gap $E_g=2.62eV$.

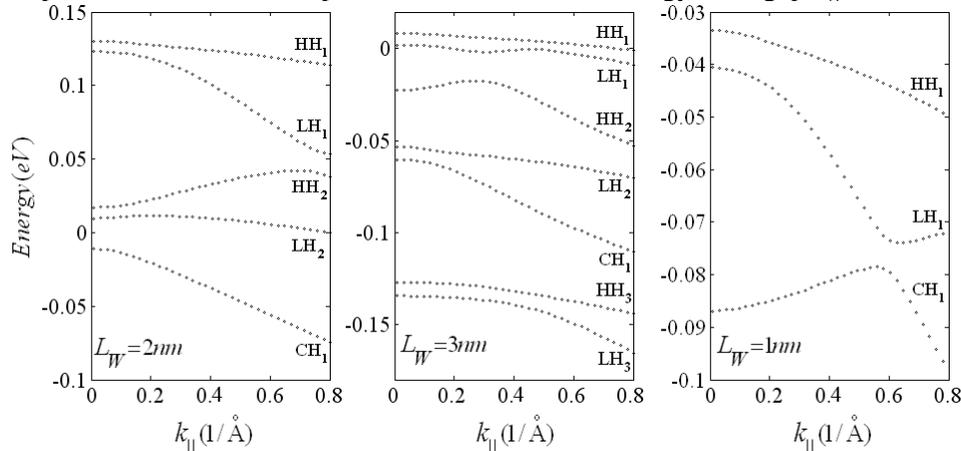

Fig. 2. Band structure for red, green and blue emitter QWs (left to right). For the red Emission $In_{0.32}Ga_{0.68}N$, for the green emission $In_{0.25}Ga_{0.75}N$ and for the blue emission $In_{0.23}Ga_{0.77}N$ was chosen for well composition. In green emission due to larger well length and weak Stark effect, subbands separation is less. The barrier material is $Al_{0.24}In_{0.38}Ga_{0.38}N$ with 16nm thickness in all emitters. All the energy subbands are plotted with respect to well material reference energy $E_v^0(In_xGa_{1-x}N)$.

To show the impact of internal field on wavefunctions profile, we illustrated the normalized envelope function corresponding to the first heavy hole energy band at $k_P=0$ in Fig. 3. Here, we assumed that the center is located at $z=5.5nm$. Deviation from symmetric situation is evident. Note that for the green emitter, the well thickness is $1nm$ greater than that of the red emitter. This may misleadingly infer that $\psi_1^{U,1}$ in green emission is turned aside much higher than in the case of red emission. An analogous deviation is credible for electron envelope functions, but in the negative direction.

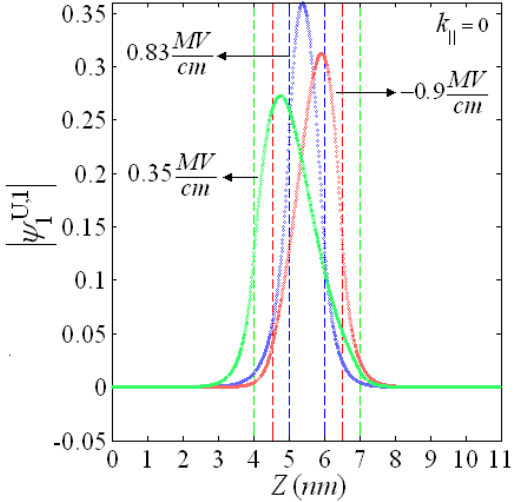

Fig. 3. Normalized envelope function $\psi_1^{U,1}$ for the first heavy hole energy band at $k_P=0$ in all there QWs. Corresponding internal fields are mentioned. The well borders are shown in each case by vertical dashed lines.

Another conclusion, which we observed during simulations, was connected to the sensitivity of material gain spectrum to carrier density. The importance rises from the requirement for setting quasi Fermi levels. One can exploit recursive schemes in order to find actual levels for both electron and hole Fermi energies. These schemes seemed to be unavoidable, as the intensity spectrum was strictly responsive to slight Fermi level deviations.

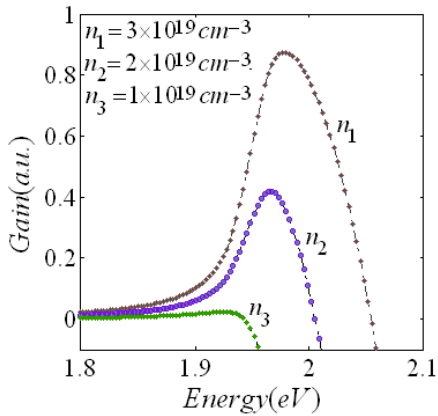

Fig. 4. Material gain in TE polarization for three various carrier densities is exhibited for the red single QW. Amplitude sensitivity to quasi Fermi levels decreases for weak carrier injection $n<n_3$.

Fig. 4 shows the material gain spectrum for TE polarization in our red single QW. The optical gain due to TE-polarized light is strongly dominant in the red QW and hence its variations reflect the total photoluminescence behavior in this emitter. It is apparent that for a minor change in carrier concentration, only from $3\times10^{19}cm^{-3}$ to $1\times10^{19}cm^{-3}$, amplitude falls within several orders of magnitude. This implies that an accurate setting is necessary for quasi Fermi levels initialization if we cross $g_m(n=n_3)$. On the other hand, to operate in spontaneous regime, injected carrier concentration must be less than $10^{19}cm^{-3}$; because for higher densities, material gain approaches the saturation condition. This ensures us that material gain in spontaneous regime is weaker than $g_m(n=n_3)$, in which sensitivity becomes less critical.

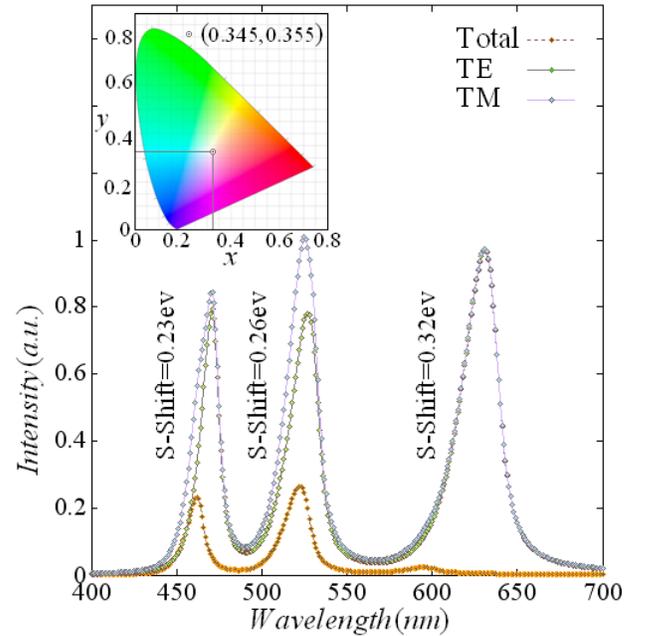

Fig. 5. Optical Intensity (photoluminescence) spectrum versus wavelength in our white LED is shown for both TE and TM-polarized light. Transitions from first conduction band to $HH_1$ and $LH_1$ affect the TE spectral peaks while transitions into $HH_2$ and $LH_2$ broaden each QW spectrum. For the blue emitter this broadening is weak as a result of distant second valence band ($HH_2$, $LH_2$). In the green and blue QWs, $CH_1$ subband is comparatively close to the settled valence quasi Fermi level and TM spectral peaks are much higher than that of red QW with a farther CH1. Inset: Chromaticity coordinates obtained for the designed MQW.

The photoluminescence intensity versus wavelength is illustrated in Fig. 5 for both TE and TM polarizations. For red and green emission ten valence bands were taken into account. In the blue case, due to considerable band splitting, just five valence bands were regarded. Carrier density was tuned in the same range for three QWs at room temperature. We observed that for carrier densities smaller than $10^{19}cm^{-3}$, TE polarization spectral peak nearly surpassed that of TM for all three QWs. This is normally arises from the fact HH and LH envelope functions chiefly contribute in the TE dipole moment matrix elements.

To achieve the total spectrum, we replaced each single spectrum with the corresponding Stokes' shift along wavelength axis and summarized over the visible range. The optical intensity spectrum exhibits three and two main peaks for TE and TM polarizations, respectively, correctly on the desired wavelengths. In all three QWs the valence quasi Fermi level is tuned close to $HH_1$ subband and therefore TE spectral peaks are mainly correspond to transitions from first conduction subband to $HH_1$ and $LH_1$. Other HH and LH subbands participate in broadening each QW spectrum. In fact, the broadening originates in more dense subbands in $k_P$ space that is not possible for the thin blue QW. Analogously, TM spectral peaks are affected by dipole moment matrix elements due to transitions into CH subband. In the red emitter, CH1 is approximately 0.15eV far from top of valence band, yielding an intense suppression for TM optical intensity.

The significant point is that the peak amplitudes have been so adjusted to exhibit nearly equal values for similar carrier injections. The chromaticity coordinates have been calculated based on the standard observer CIE 1931 color matching functions [40]. This resulted in the coordinates of $(x,y)=(0.355,0.345)$, very close to the $(0.33,0.33)$ coordinates of ideal white light with flat spectrum, as illustrated in the inset of Fig. 5. In contrast to the existing white LEDs which emit 'cold' light with strong blue emission, our proposed device is expected to emit 'warm' white light, slightly resembling the yellowish appearance of sunlight.

## 8. Conclusion

In this paper, we employed $k.p$ transfer matrix method to analyze valence band structure and relevant envelope functions in a GaN-based heterostructure. Basis set was chosen in a way that corresponding envelope functions could simplify optical gain expression. The dipole moment matrix elements were derived based on envelope functions both in TE and TM polarizations. Then, we considered the internal field and Stark effect on correlation terms in gain expression. The effect focused on envelope functions and carrier distributions deviation from the symmetric condition. Two major parameters were examined to adjust the best intensity and proper band gap with the aid of internal field: biaxial strain and well thickness. While strain altered piezoelectric field, well thickness enhanced or reduced sloped potential influence. Combining these effects gave a band gap widening or narrowing shift in energy eigenvalues. On the other hand, Stark effect separates electron/hole distributions especially when the QW depth is comparable to $\Delta E_g$. With the quaternary composition of barrier, $Al_xIn_yGa_{1-x-y}N$, we succeeded to enlarge $\Delta E_g$ with respect to $E_i \times L_W$ and avoid an intense separation between electron/hole wavefunctions. It is noteworthy to mention that excitonic localization effect aids us achieving a redshift to higher wavelengths without increasing indium content higher than $x=0.32$ inside red emitter. Finally an optimized spectrum was obtained which from its chromaticity coordinates we can expect a warm white light emission from the proposed LED.

## Appendix

Here, the band parameters are listed, which are used in band structure calculation of wurtzite GaN-family QWs.

Table I
Band parameters for GaN, InN, and AlN

| Parameter | GaN | InN | AlN |
|---|---|---|---|
| Energy band gap (eV) | 3.44[a] | 0.76[a] | 6.28[a] |
| Relative dielectric constant | 9.4 | 15 | 8.5 |
| Lattice constant $a$ ($nm$) | 3.19 | 3.54 | 3.11 |
| $\gamma_{1P}$ | 2.38[f] | 3.2[f] | 1.55[f] |
| $\gamma_{1z}$ | 2.87[f] | 3.69[f] | 1.8[f] |
| $\gamma_{2P}$ | 0.78[f] | 1.19[f] | 0.64[f] |
| $\gamma_{2z}$ | 0.71[f] | 1.13[f] | 0.44[f] |
| $\gamma_{3P}$ | 0.93[f] | 1.31[f] | 0.54[f] |
| $\gamma_{3z}$ | 1.09[f] | 1.49[f] | 0.65[f] |
| Hydrostatic deformation potential $V_a$ (eV) | -6.12[b] | -4.1[c] | - |
| Shear deformation potential $V_b$ (eV) | -2.34[b] | -2.1[b] | - |
| Crystal field splitting energy $\Delta_{cr}$ (eV) | 0.023[d] | 0.037[d] | -0.06[d] |
| Spin-orbit splitting energy $\Delta_{so}$ (eV) | 0.016[d] | 0.011[d] | - |
| Spontaneous polarization factor $P_{sp}$ (C/m$^2$) | -0.034[e] | -0.042[e] | -0.09[e] |
| Piezoelectric constant $d_{31}$ ($pm$/V) | -1.6[b] | -2.6[b] | - |

[a] Ref. [21]  [b] Ref. [24]
[c] Ref. [37]  [d] Ref. [13]
[e] Ref. [38]  [f] Ref. [39]

**Acknowledgement**
All computations were done using MATLAB codes developed by Milad Khoshnegar (email: milad.khoshnegar@gmail.com) at Sharif University of Technology. The authors wish to thank Iranian National Science Foundation (INSF) for partial support of this work. The authors also gratefully acknowledge the valuable comments of anonymous reviewers of this work.